# SDSS J162718.39+120435.0 - a dwarf nova in the period gap

Jeremy Shears, Steve Brady, Boris Gänsicke, Tom Krajci, Ian Miller, Yenal Ögmen, Jochen Pietz and Bart Staels

**Abstract**

SDSS J162718.39+120435.0 was suspected of being a dwarf nova from spectroscopic observations made available by the Sloan Digital Sky Survey. Photometry conducted during the 2008 outburst, the first ever outburst of this object detected and followed up in real time, shows that the outburst reached magnitude 14.6 at maximum, had an amplitude of 4.6 magnitudes and lasted for at least 18 days. Common superhumps were detected with an amplitude of up to 0.4 magnitudes, confirming it to be a member of the SU UMa family. Initially the superhump period, $P_{sh}$, was 0.10993(7) days (2.638 hours), but it subsequently reduced to $P_{sh}$ = 0.10890(9) days (2.614 hours) later in the outburst. The period change corresponds to the end of the plateau period of the outburst. The orbital period, $P_{orb}$, was estimated from the two $P_{sh}$ values as between 0.1026 d and 0.1016 d, which places SDSS1627 near the centre of the period gap in the orbital period distribution of cataclysmic variables.

**Introduction**

Dwarf novae are a class of cataclysmic variable (CV) star in which a white dwarf primary accretes material from a secondary star via Roche lobe overflow. The secondary is usually a late-type main-sequence star. In the absence of a significant white dwarf magnetic field, material from the secondary passes through an accretion disc before settling on the surface of the white dwarf. As material builds up in the disc, a thermal instability is triggered that drives the disc into a hotter, brighter state causing an outburst in which the star apparently brightens by several magnitudes [1]. Dwarf novae of the SU UMa family occasionally exhibit superoutbursts which last several times longer than normal outbursts and may be up to a magnitude brighter. During a superoutburst the light curve of a SU UMa system is characterised by superhumps. These are modulations in the light curve which are a few percent longer than the orbital period. They are thought to arise from the interaction of the secondary star orbit with a slowly precessing eccentric accretion disc. The eccentricity of the disc arises because a 3:1 resonance occurs between the secondary star orbit and the motion of matter in the outer accretion disc. For a more detailed review of SU UMa dwarf novae and superhumps, the reader is directed to references 1 and 2.

The Sloan Digital Sky Survey ("SDSS") is conducting a major sky survey using a dedicated 2.5 m telescope at Apache Point, New Mexico, USA, equipped with a 120 mega-pixel camera and a pair of spectrographs that can obtain spectra of hundreds of objects simultaneously. We have identified SDSS J162718.39+120435.0 (hereafter "SDSS1627") as a dwarf nova candidate in the spectroscopic data release 6 (DR6) of the SDSS [3]. The spectrum of SDSS1627 (Figure 1) is characterised by strong Balmer and HeI emission lines, plus weak emission of HeII and/or the NIII/CIII Bowen blend. No spectroscopic signature of either the white dwarf or the companion star are detected. Overall, the spectrum is typical of a SU UMa type dwarf nova. In quiescence it has g= 19.2 and r = 19.1. Following the discovery of the outburst discussed in this paper, Patrick Wils noted that the object was recorded in outburst on the POSS 1 blue and red plates from 1955 Mar, at magnitude B = 14.71 and R = 14.67 [4]. We are not aware of any other outbursts of this object. SDSS 1627 is located in Hercules at RA 16 h 27 min 18.39 sec Dec +12 deg 04 min 35.0 sec (J2000)

**The 2008 superoutburst**

The 2008 outburst of SDSS 1627 was detected on May 30.093 at 15.76C by SB [5] and is the first outburst to be detected and followed up in real time. Time series photometry was begun the following night and continued during 33 separate sessions over the following 14 days yielding 154 hours of data. Table 1 summarises the instrumentation used and Table 2 contains a log of the time-series runs. In all cases raw images were flat-fielded and dark-subtracted, before being analysed using commercially available differential aperture photometry software relative to comparison stars with V-band photometry given by Henden [6]. An image of the star in outburst is shown in Figure 2. In the paper we shall frequently refer to dates in the truncated form JD = JD – 2454000.

The overall light curve of the outburst is shown in the top panel of Figure 3. It appears that the outburst was caught in its early stages on JD616 since it brightened by 1 magnitude in the following 24 h. The next 4 days (JD617 to 621) appear to correspond to the plateau phase commonly seen near the beginning of dwarf nova outbursts when the star reached ~14.6C (C = CCD, no filter). The star then showed an approximately linear decline at 0.17 mag/d from JD 623 to 629, typical of a dwarf nova in decline. Finally there was a slightly sharper decline between JD629 and 631 and by JD 634.4, the star was at 19.1C close to its pre-outburst magnitude of 19.3C on JD 587. Thus the outburst lasted at least 18 days and the amplitude was 4.6 magnitudes. Further modest brightness variations were detected after this. For example on JD 640.5 the star reached 17.1C and on JD 642 the star was found to be varying between 18.3 and 19.0C. None of these appear to be the post outburst rebrightening events, sometimes called echo outbursts, that have seen in some very short orbital period systems of the WZ Sge type, as they were neither as sustained nor as bright. Brightness variations during quiescence, similar to those seen here, are common in dwarf novae.

Expanded plots of the photometry data are shown in Figures 4a to c, all drawn to the same magnitude and time scale. Modulations were apparent on the first night of photometry, but they did not appear to have a regular structure and may have been developing superhumps. However, by the following night common superhumps having peak-to-peak amplitude of ~0.2 mag were obvious, confirming this to be a superoutburst on an SU UMa type dwarf nova. The superhumps appeared to be growing and by the next night (JD619) they reached their greatest amplitude of ~0.4 mag. As the outburst continued, the superhumps diminished in amplitude to ~0.3 mag, subsequently regrowing to ~0.38 mag towards the end of the outburst.

**Measurement of the superhump period**

To study the superhump behaviour, we first extracted the times of each sufficiently well-defined superhump maximum from the individual light curves according to the Kwee and van Woerden method [7] using the Peranso software [8]. Times of 50 superhump maxima were found and are listed in Table 3. Following a preliminary assignment of superhump cycle numbers to these maxima, an analysis of the times of maximum allowed us to obtain the following linear superhump maximum ephemeris:

$$JD_{Max} = 2454618.7875(33) + 0.10935(11) \times E$$

The observed minus calculated (O–C) residuals relative to this ephemeris are shown in Table 3 and plotted in the bottom panel of Figure 3. This shows that the period was not constant during the outburst. Putting aside the first two superhumps, which occurred during the development of the superhumps, there appear to be two superhump regimes in operation during the outburst, each having a discrete superhump period, $P_{sh}$:

JD619 to 621        $P_{sh} = 0.10993(7)$ d (2.638 h)
JD623 to 630        $P_{sh} = 0.10890(9)$ d (2.614 h)

Therefore sometime between JD 621 and 623 (cycle 28 and 45) the superhump period changed. This appears to correspond to the end of the plateau phase and the beginning of the decline. A reduction in $P_{sh}$ has been observed in many SU UMa stars as the outburst progresses, including DV UMa, IY UMa and V452 Cas [9, 10, 11] and may be explained by the accretion disc emptying and shrinking. Usually the reduction in $P_{sh}$ is continuous, although in the case of V452 Cas it does seem to have been more abrupt [11] with two discrete values of $P_{sh}$, similar to what we have observed in SDSS 1627.

To confirm our measurement of $P_{sh}$, we carried out a period analysis of the combined data from the plateau phase between JD 619 and 621 using the Data Compensated Discrete Fourier Transform (DCDFT) algorithm in Peranso, after subtracting the mean and linear trend from each of the light curves. This gave the power spectrum in Figure 5 which has its highest peak at a period of 0.1099(7) d (and its 1d aliases), which we interpret as $P_{sh}$; this value is consistent with our earlier measurement. The superhump period error estimate is derived using the Schwarzenberg-Czerny method [12]. Several other statistical algorithms in Peranso gave the same value of $P_{sh}$. In a similar manner we analysed the data from JD 623 to 630, which yielded $P_{sh} = 0.1089(4)$ d (and its 1 day aliases; power spectrum not shown), which is again consistent with the result from analysing times of superhump maxima.

Removing $P_{sh}$ from each of the two power spectra leaves only weak signals, none of which has any significant relationship to the superhump periods. We searched for signals that could be associated with an orbital hump, and which could therefore yield the hitherto unknown orbital period, but none were forthcoming. Phase diagrams of the data from each section of the superhump regime, folded on their respective values of $P_{sh}$, are shown in Figure 6 and 7, where two cycles are shown for clarity. This confirms that $P_{sh}$ remained constant during each of the two regimes, but changed sometime in between.

**Discussion**

The values for $P_{sh}$ derived above suggest that the orbital period, $P_{orb}$, of SDSS 1627 is relatively long for a SU UMa type dwarf novae. The $P_{sh}$ values of SDSS 1627 are similar to that of TU Men, which has $P_{sh} = 0.1256$ d and a superhump period excess, $\varepsilon$, of 0.0717, where $\varepsilon = (P_{sh} - P_{orb}) / P_{orb}$ [13]. Assuming SDSS 1627 has a similar $\varepsilon$, we estimate its $P_{orb}$ is between 0.1026 d and 0.1016 d (2.46 and 2.44 h respectively), based on the two values of $P_{sh}$. Since the value of $\varepsilon$ depends on the mass ratio of the binary stars, which is presently unknown for SDSS 1627, our $P_{orb}$ values are only approximate. Radial velocity measurements or photometry at quiescence are required to measure $P_{orb}$ accurately.

Figure 8 shows the orbital period distribution of CVs and dwarf novae using data taken from version 7.9 of the "Ritter and Kolb" catalogue [14]; objects of uncertain period (those marked with a colon in ref 14) were excluded. In the case of SDSS 1627 both estimates of $P_{orb}$ place it near the centre of the so-called period gap of 2.15 to 3.18 h in the period distribution of CVs [15] shown in Figure 8, where the ends of the red bar indicate the two values of $P_{orb}$ of estimated in this study. The period gap represents a division between CVs of short orbital period and low mass-transfer rates (below the gap) and long orbital period high mass-transfer systems that are mainly driven by magnetic breaking (above the gap) [1, 16]. It is generally assumed that systems in the period gap have formed there, i.e. they filled their Roche lobe at an orbital period close to the present one and that their secondary stars have yet to undergo the structural changes of secondary stars that have occurred in systems outside the period gap [16]. Detailed spectroscopic studies of gap-born secondary stars are few partly because rather few systems exist and partly because of the technical difficulties in isolating the secondary component from other light sources in the binary system. In this regard, SDSS 1627 may present an opportunity to further these studies.

**Conclusions**

Time resolved photometry during the 2008 outburst of SDSS 1627 showed that the star reached magnitude 14.6 at maximum and had an amplitude of 4.6 magnitudes. The outburst lasted at least 18 days. After the initial rise to full outburst, the next four days corresponded to the plateau phase and this was followed by an approximately linear decline at 0.17 mag/d. After the main outburst was over, further low level brightness variations were detected.

Common superhumps with a peak-to-peak amplitude of ~0.2 mag were visible early in the outburst, showing that SDSS 1627 is a member of the SU UMa family of dwarf novae. These grew to ~0.4 mag, subsequently diminishing in amplitude to ~0.3 mag and finally regrowing to ~0.38 mag towards the end of the outburst.

During the early part of the outburst the superhump period $P_{sh} = 0.10993(7)$ d, subsequently decreasing to $P_{sh} = 0.10890(9)$ d. The period change corresponds to the end of the plateau phase of the outburst and the beginning of the decline. The orbital period, $P_{orb}$, was estimated from the two $P_{sh}$ values as between 0.1026 d and 0.1016 d, which places SDSS1627 near the centre of the period gap in the period distribution of CVs.

We encourage observers, both visual and CCD-equipped, to monitor this star for future outbursts with the aim of establishing the outburst frequency and supercycle length. Such monitoring may also shed light on the variations in brightness observed during quiescence.

**Acknowledgements**

The authors thank Dr. Arne Henden for making his comparison star photometry available to us. We acknowledge the use of SIMBAD, operated through the Centre de Données Astronomiques (Strasbourg, France) and the NASA/Smithsonian Astrophysics Data System. Finally, we thank our referees for suggestions which have improved the paper.

**Addresses:**
JS: "Pemberton", School Lane, Bunbury, Tarporley, Cheshire, CW6 9NR, UK [bunburyobservatory@hotmail.com]
SB: 5 Melba Drive, Hudson, NH 03051, USA [sbrady10@verizon.net]

BG: Department of Physics, University of Warwick, Coventry, CV4 7AL
[Boris.Gaensicke@warwick.ac.uk]
TK: CBA New Mexico, PO Box 1351 Cloudcroft, New Mexico 88317, USA
[tom_krajci@tularosa.net]
IM: Furzehill House, Ilston, Swansea, SA2 7LE, UK [furzehillobservatory@hotmail.com]
YÖ: Geçitkale, Magosa, via Mersin , North Cyprus [Yenalogmen@yahoo.com]
JP: Nollenweg 6, 65510 Idstein, Germany [j.pietz@arcor.de]
BS: Alan Guth Observatory, Koningshofbaan 51, Hofstade, Aalst, Belgium
[staels.bart.bvba@pandora.be]

| Observer | Telescope | CCD (unfiltered) |
|---|---|---|
| JS | 0.28 m SCT | Starlight Xpress SXVF-H9 |
| SB | 0.4 m reflector | SBIG ST-8XME |
| TK | 0.28 m SCT | SBIG ST-7E CCD |
| IM | 0.35 m SCT | Starlight Xpress SXVF-H16 |
| YÖ | 0.36 m SCT | Meade DSI Pro II |
| JP | 0.28 m SCT | SBIG ST-8XME |
| BS | 0.28 m SCT | Starlight Xpress MX716 |

**Table 1: Equipment used**

| Date in 2008 (UT) | Start time (JD-2454000) | Duration (h) | Observer |
|---|---|---|---|
| May 31 | 617.63 | 7.9 | TK |
| May 31 | 618.42 | 3.6 | JP |
| Jun 1 | 618.64 | 7.9 | TK |
| Jun 1 | 619.42 | 3.6 | IM |
| Jun 2 | 619.59 | 4.8 | SB |
| Jun 2 | 619.62 | 8.2 | TK |
| Jun 3 | 620.59 | 7.2 | SB |
| Jun 3 | 620.62 | 8.2 | TK |
| Jun 3 | 621.27 | 2.6 | YÖ |
| Jun 3 | 621.38 | 3.1 | JP |
| Jun 3 | 621.43 | 4.8 | IM |
| Jun 4 | 621.63 | 7.9 | TK |
| Jun 5 | 623.41 | 3.4 | JS |
| Jun 5 | 623.44 | 1.0 | IM |
| Jun 6 | 623.63 | 7.9 | TK |
| Jun 6 | 624.26 | 3.4 | YÖ |
| Jun 6 | 624.41 | 4.1 | JS |
| Jun 6 | 624.46 | 3.1 | IM |
| Jun 7 | 625.41 | 4.1 | JS |
| Jun 7 | 625.43 | 2.6 | IM |
| Jun 8 | 625.59 | 1.4 | SB |
| Jun 8 | 626.40 | 3.1 | JP |
| Jun 8 | 626.44 | 3.4 | IM |
| Jun 9 | 627.43 | 3.6 | IM |
| Jun 10 | 628.38 | 4.1 | BS |
| Jun 10 | 628.38 | 3.1 | JP |
| Jun 10 | 628.44 | 2.9 | IM |
| Jun 12 | 629.59 | 4.3 | SB |
| Jun 12 | 629,62 | 8.2 | TK |
| Jun 12 | 630.44 | 3.4 | IM |
| Jun 13 | 630.59 | 4.3 | SB |
| Jun 13 | 630.63 | 7.7 | TK |
| Jun 14 | 631.63 | 4.8 | TK |

**Table 2: Log of time-series observations**

| Superhump cycle number | Time of maximum (JD-2454000) | O-C (cycles) |
|---|---|---|
| 0 | 618.76640 | -0.08916 |
| 1 | 618.87348 | -0.10992 |
| 6 | 619.44156 | 0.08514 |
| 7 | 619.55165 | 0.09191 |
| 8 | 619.66318 | 0.11184 |
| 8 | 619.66189 | 0.10005 |
| 9 | 619.76901 | 0.07965 |
| 9 | 619.76895 | 0.07910 |
| 10 | 619.88230 | 0.11568 |
| 17 | 620.65223 | 0.15665 |
| 17 | 620.65198 | 0.15437 |
| 18 | 620.75700 | 0.11477 |
| 18 | 620.75869 | 0.13022 |
| 19 | 620.86929 | 0.14166 |
| 23 | 621.30691 | 0.14367 |
| 24 | 621.42222 | 0.19817 |
| 25 | 621.53016 | 0.18528 |
| 26 | 621.64278 | 0.21518 |
| 27 | 621.74956 | 0.19168 |
| 28 | 621.85633 | 0.16808 |
| 45 | 623.71159 | 0.13434 |
| 46 | 623.81903 | 0.11687 |
| 47 | 623.92564 | 0.09182 |
| 51 | 624.36264 | 0.08816 |
| 52 | 624.47578 | 0.12282 |
| 52 | 624.47508 | 0.11642 |
| 53 | 624.58004 | 0.07627 |
| 61 | 625.45187 | 0.04911 |
| 61 | 625.45019 | 0.03374 |
| 62 | 625.56260 | 0.06173 |
| 70 | 626.44007 | 0.08615 |
| 71 | 626.54286 | 0.02615 |
| 80 | 627.52094 | -0.02932 |
| 88 | 628.38919 | -0.08925 |
| 88 | 628.39013 | -0.08066 |
| 89 | 628.49728 | -0.10078 |
| 89 | 628.49911 | -0.08404 |
| 89 | 628.49874 | -0.08743 |
| 100 | 629.70012 | -0.10087 |
| 100 | 629.69843 | -0.11632 |
| 101 | 629.80658 | -0.12730 |
| 102 | 629.91539 | -0.13224 |
| 107 | 630.46359 | -0.11898 |
| 108 | 630.57105 | -0.13626 |
| 109 | 630.67766 | -0.16132 |
| 109 | 630.68187 | -0.12282 |
| 110 | 630.78603 | -0.17028 |
| 111 | 630.89500 | -0.17375 |
| 118 | 631.65925 | -0.18473 |
| 119 | 631.76894 | -0.18161 |

**Table 3: Times of superhump maximum**

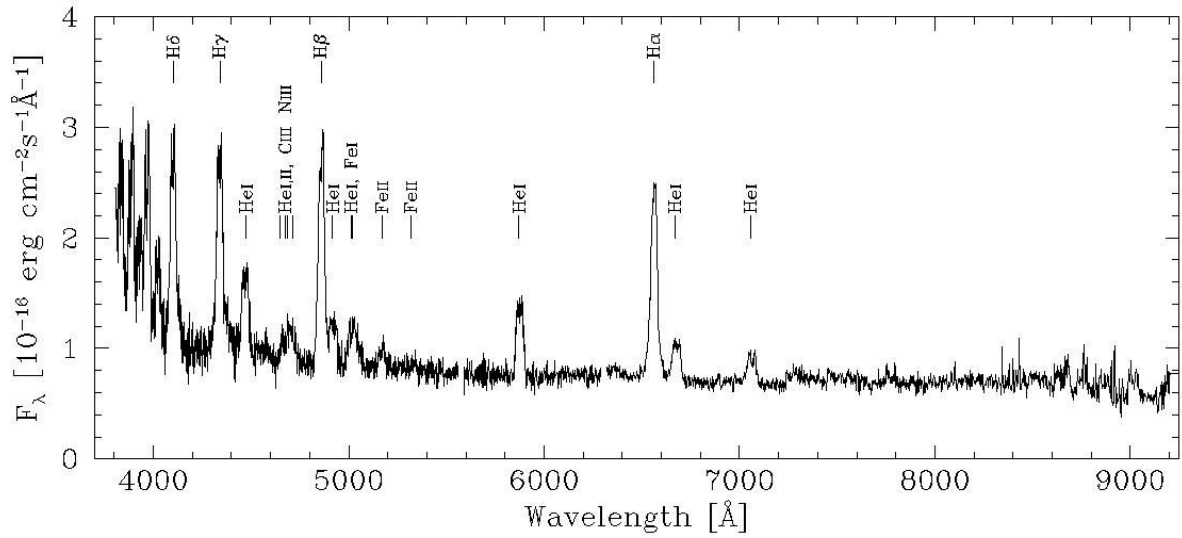

**Figure 1: Spectrum of SDSS1627**
2005 Jul 29, 2.5 m reflector. Data from ref. 3

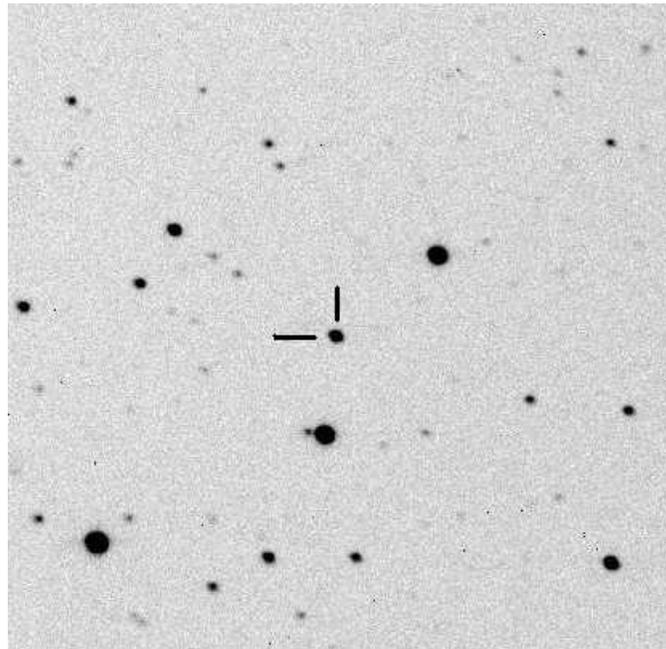

**Figure 2: SDSS1627 in outburst**
2008 Jun 2. Field 7' x 7' with North at top, East to left. See Table 1 for instrumentation
(*Steve Brady*)

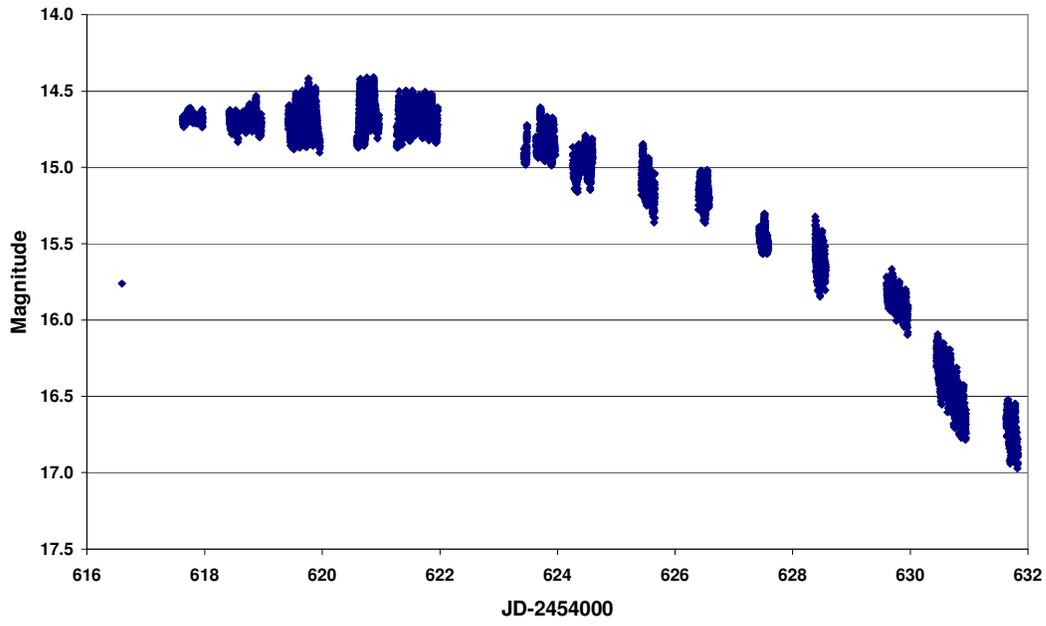

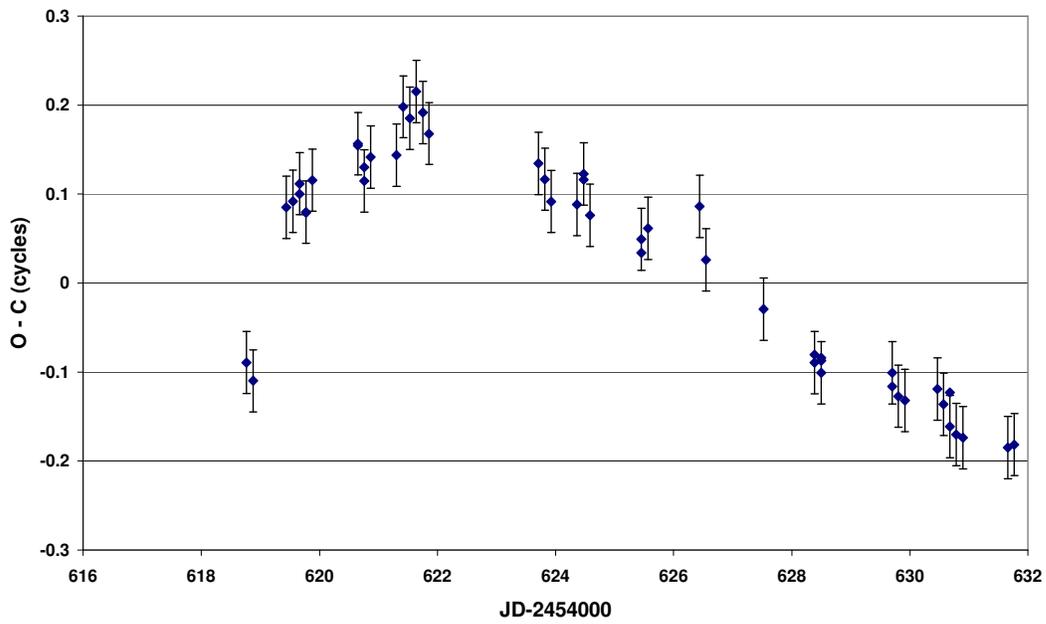

**Figure 3: Unfiltered CCD light curve of the outburst (top) and O-C diagram of superhump maxima (bottom)**

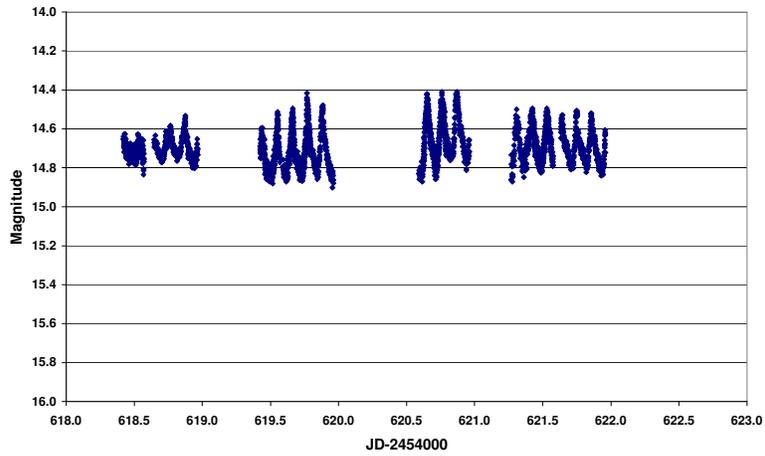

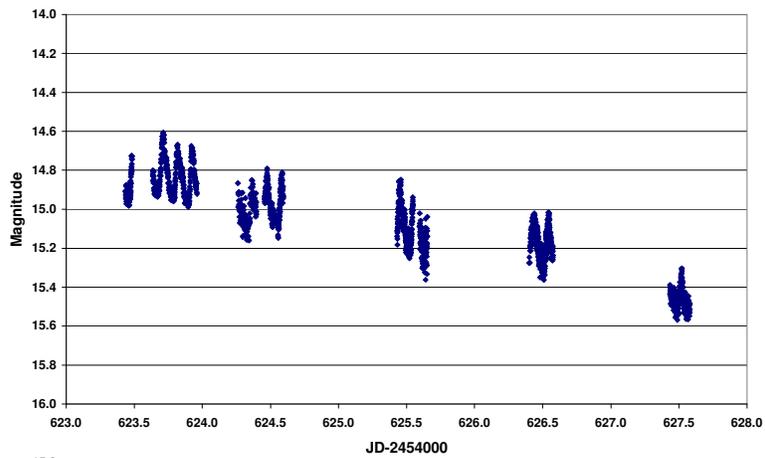

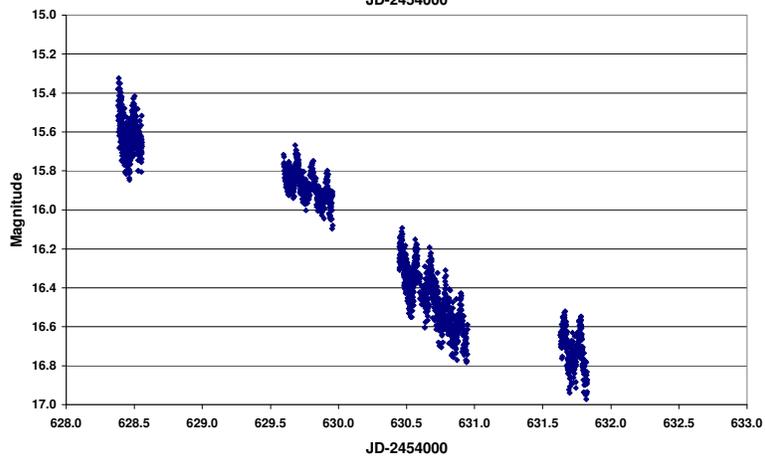

**Figure 4: Time series photometry data, unfiltered CCD**
(a) top, (b) middle, (c) bottom

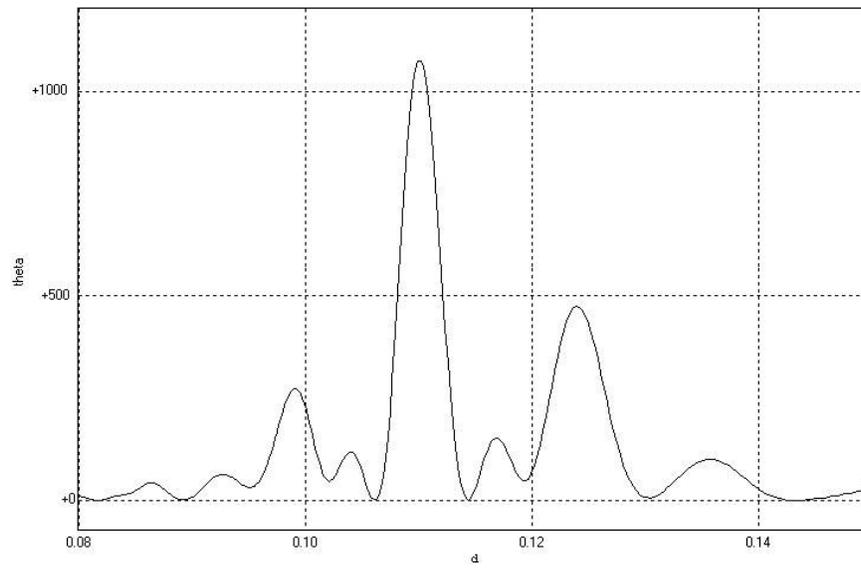

**Figure 5: Power spectrum of combined time-series data from JD619 to 621**

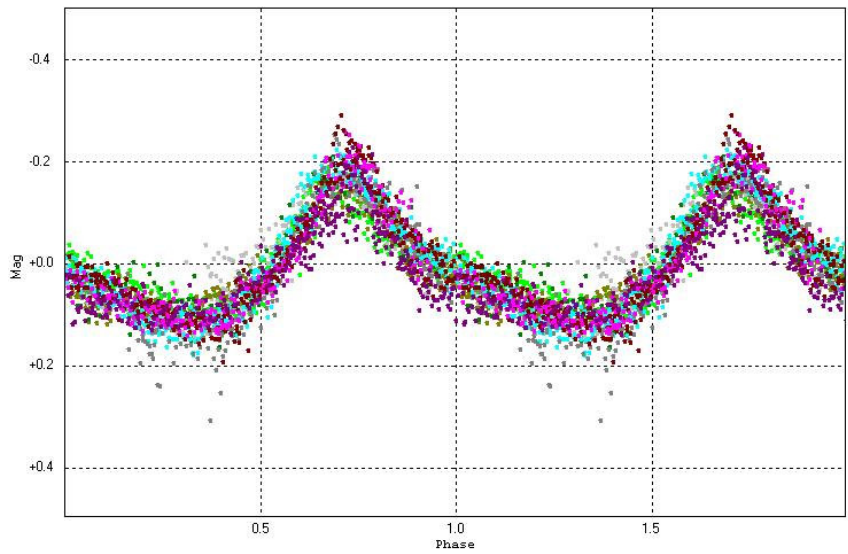

**Figure 6: Phase diagram of data from JD619 to 621 folded on P$_{sh}$ of 0.1099 d**

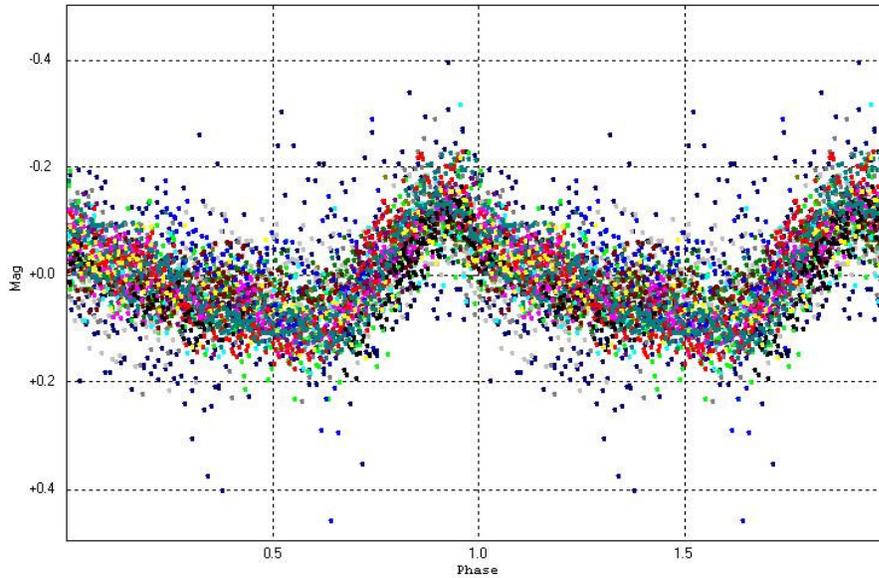

**Figure 7: Phase diagram of data from JD 623 to 630 folded on P$_{sh}$ of 0.1089 d**

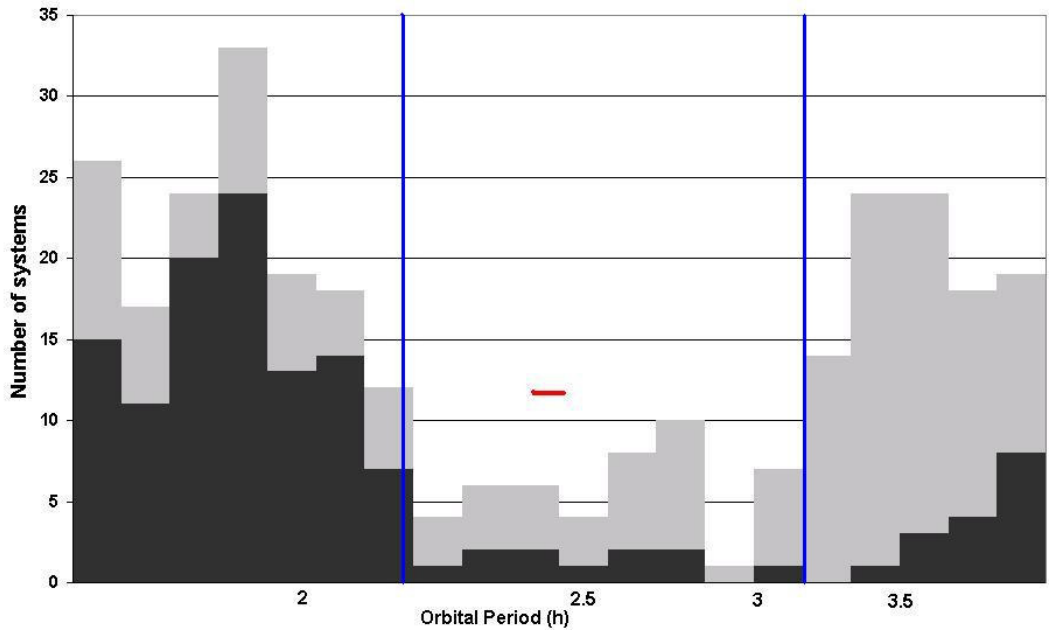

**Figure 8: Period distribution of CVs showing the period gap**
Light grey boxes represent all CV sub-types and dark boxes are dwarf novae only. Vertical blue lines represent the extent of the period gap given in ref 15. The red bar indicates the position of SDSS 1627, which is not included in the histograms.